\documentclass
[pre,showpacs,
tightenlines,
aps,
secnumarabic,
amsmath,amssymb,floatfix]{revtex4}
\usepackage{docs}

\newif\ifpdf
\ifx\pdfoutput\undefined
\pdffalse 
\else
\pdfoutput=1 
\pdftrue
\fi

\ifpdf
\usepackage[pdftex]{graphicx} 
\else
\usepackage{graphicx}
\fi

\DeclareGraphicsExtensions{.eps, .jpg,.epsc,.gif} 
\normalsize
\begin{document}
\title{
An Information-Theoretic Approach to Network Modularity
}
\author{Etay Ziv$^{\dagger}$}
\affiliation{College of Physicians \& Surgeons,\\
Department of Biomedical Engineering,\\
Columbia University}
\author{Manuel Middendorf$^{\dagger}$}
\affiliation{Department of Physics,\\
Columbia University}
\author{Chris Wiggins}
\affiliation{Department of Applied Physics and Applied Mathematics,\\
Center for Computational Biology and Bioinformatics,\\ 
Columbia University \\
$^\dagger${\textup{These authors contributed equally to this work.}}
}

\begin{abstract}
Exploiting recent
developments in information theory, 
we propose, illustrate, and validate a principled
information-theoretic algorithm for module discovery and resulting
measure of network modularity.
This measure is
an order parameter (a dimensionless number between $0$ and $1$).
Comparison is made
to other approaches to module-discovery
and to quantifying network modularity using Monte Carlo generated Erd\"os-like modular networks.
Finally, the Network Information Bottleneck (NIB) algorithm is applied to
a number of real world networks, including the ``social" network
of coauthors at the APS March Meeting 2004.
\end{abstract}

\maketitle

\section{Introduction}
A goal of modeling is to describe the system in terms of less complex degrees of freedom while retaining information deemed relevant~\cite{bottleneck}. Approaches to complexity reduction for networks include characterizing the network in terms of simple statistics (such as degree distributions ~\cite{barabasi_sf} or clustering coefficients~\cite{small_world}), subgraphs over-represented relative to an assumed null model (see for example,~\cite{HandL, alon02b, Lewistone}), and communities (see for example,~\cite{newman}). 

In the case of communities, networks are coarse-grained into clusters of nodes, or modules, where nodes belonging to one cluster are highly interconnected, yet have relatively few connections to nodes in other clusters. This type of network complexity reduction may be particularly promising as an approach to network analysis, since many naturally-occurring networks, including biological~\cite{modular:murray:nature} and sociological~\cite{newman, WassermanFaust} networks are thought to be modular. Clearer, quantitative understanding of these ideas 
would be valuable in finding reduced complexity
descriptions of networks, in visualizing networks, and in 
revealing global design principles. Two current challenges facing the community regarding network modularity include (i) the ability to quantify to what extent a given network is ``modular" and (ii) the ability to identify the modules of a given network. 

With regard to quantifying modularity, to our knowledge no mathematical definition has yet been proposed for a measure of modularity that could compare networks regardless of size, origin, or choice of partitioning. In their recent book, Schlosser and Wagner \cite{wagner} write ``a generally accepted definition of a module does not exist and different authors use the concept in quite different ways." They proceed to warn of the ``danger that modularity will degenerate into a fashionable but empty phrase unless its precise meaning is specified."
Some steps in this 
direction have been suggested by Newman's ``assortativity coefficient"~\cite{newman_assort}, which quantifies the level of 
assortative mixing in a network, and its unnormalized
form, called ``modularity"~\cite{newman}. 
However, these measures quantify the quality of a particular partitioning of the network for a given
number of modules, but are {\em not} a property of the network itself that could
serve to compare networks of different origins. 

As for module discovery, a range of techniques for identifying the modules in a 
network have been utilized with various success. In his review article, Newman~\cite{newman_review} summarizes these efforts under the broad category of hierarchical clustering in which one poses a similarity metric between pairs of vertices. Hierarchical clustering can be agglomerative, where the most similar nodes are iteratively merged (\emph{e.g.}~\cite{barabasi}) or divisive, where edges between the least similar nodes are iteratively removed (\emph{e.g.}~\cite{newman}).  
By modifying traditional divisive approaches to focus on most ``between" edges rather 
than least similar vertices, Newman and Girvan~\cite{newman} recently 
proposed a new class of divisive algorithms for finding modules.
Various measures of ``edge-betweenness" are defined
to identify edges that lie between communities rather than within 
communities. By iteratively removing edges with highest betweenness
one can break down the network into disconnected components which define
the modules. 

In this manuscript we take a markedly different approach. The problem of finding reduced descriptions of systems while retaining information deemed relevant has been well-studied in the learning theory community. In particular, the information bottleneck~\cite{bottleneck,slonimthesis} 
provides a 
unified and principled framework for complexity reduction.
By applying the information bottleneck on probability distributions
defined by graph diffusion, we propose a new, principled,
information-theoretic algorithm to identify modules
in networks. 
We demonstrate that the Network Information Bottleneck (NIB) algorithm 
 outperforms the currently used technique of edge-betweenness 
 (i) in correctly assigning nodes to modules 
and (ii) in determining the optimal number of existing modules.
Moreover, the new method naturally defines a network modularity measure which
can compare any two undirected networks to the extent to which the
topology of each can be summarized by modules over all scales. Information-theoretic bounds constrain this measure to be between $0$ and $1$. 
Finally, we apply our method to a collaboration network derived from the APS March Meeting 2004 and the \emph{E. coli} genetic regulatory network.
 
\section{The Information Bottleneck: A Review}
Brief~\cite{bottleneck} and detailed
\cite{slonimthesis} discussions of the information bottleneck can be found
elsewhere; we here review only the most salient features. The fundamental quantity in information theory is Shannon entropy 
$H[p(x)]\equiv -\sum_xp(x)\log p(x)$ measuring lack
of information (or disorder) in a random variable $X$, and uniquely 
(up to a constant) 
defined by three plausible axioms~\cite{shannon}.
Knowledge of a second random variable $Y$ decreases the
entropy in $X$ on average by an amount
\begin{equation}
I(X,Y) \equiv H(X)-\langle H(X|Y)\rangle\equiv
H[p(x)] - \sum_yp(y)H[p(x|y)]
\label{eqt:mi}
\end{equation}
called the {\em mutual information}~\cite{cover-thomas-book}, the average 
information gained about $X$ by the knowledge of $Y$. 
Eq. (\ref{eqt:mi}) is equivalent to 
\begin{equation}
I(X,Y)    =  \sum_x \sum_y p(x,y) \log \frac{p(x,y)}{p(x)p(y)}
=   \langle \log \frac{p(x,y)}{p(x)p(y)}\rangle
\end{equation}
revealing its symmetry in $X$ and $Y$. The mutual information thus measures
how much information one random variable tells about the other, 
and is the basis of the information bottleneck.

Clustering can generally be described as the problem of extracting 
a compressed description of some data that captures information 
deemed relevant or meaningful. For example, 
we might want to cluster protein sequences, expecting that the cluster assignments 
contain information about the 
fold
of the proteins; 
or we might want to cluster words in documents, expecting that the 
clusters capture information about the topic in which the words appear. 
Tishby et al.'s~\cite{bottleneck} key insight into this problem 
is the inclusion in the clustering algorithm of
another random variable, called 
the \emph{relevance variable}, 
which describes the information to be preserved.
In the case of protein sequences, the relevance variable 
might be the protein
fold;
in the case of clustering words over documents, the relevance variable might be the topic. 

Let $x\in X$ be the input random variable (e.g., protein sequences 
in the set of all observed sequnces; 
or words in a given dictionary, in the two examples above), $y\in Y$ the relevance variable, 
and $z\in Z$ the cluster assignment random variable~\footnote{We use $Z$ here 
as the cluster variable rather than $T$ as in many papers~\cite{slonimthesis} 
in order to avoid confusion with time $t$ and temperature $T$, which will appear later.}. 
The information bottleneck outputs 
a probabilistic cluster assignment function $p(z|x)$ equal to
the probability to be in cluster $z$ for a given input $x$. 
The clustering minimizes the mutual information between $X$ and $Z$ 
(``maximally
compressing the data set''),
while constraining the possible loss in mutual information between $Z$ and $Y$
(``preserving relevant information'').
In other words, one seeks to pass or 
squeeze the information that $X$ provides about $Y$ through the 
``bottleneck'' formed by the compressed $Z$. 

The simplicity of the model Z relative to that of the world X
is quantified by the entropy reduction 
${\cal S}[p(z|x)]\equiv 
	H(X)-I(X,Z)=
	H(X,Z)-H(Z)>0$.
The gain in simplicity, however, comes with a loss of fidelity
in our description of the world, quantified by the error
${\cal E}\equiv I(X,Y)-I(X,Z)>0$,
the
loss in information about the world when described by a model Z
instead of the primitive description X.
The trade-off between the error and the simplicity can
be expressed in terms of the functional 
\begin{equation}
{\cal F}[p(z|x)]={\cal E}-T{\cal S}={\cal F}_0-I(X,Z)+TI(X,Z)
\label{eqn:free_energy}
\end{equation}
in which the temperature T parameterizes the relative importance of 
simplicity over fidelity.
The term ${\cal F}_0$ is independent
of the cluster assignment $p(z|x)$. Since $p(y|x,z)=p(y|x)$, this
is the only degree of freedom over which the free energy ${\cal F}$
is to be minimized.
In the annealed ground state ($T\rightarrow 0$) 
each possible state of the world $x\in X$ is
assigned with unit probability to 
one and only one state of the model $z\in Z$
(i.e., $p(z|x)\in\{0,1\}$, a limit called "hard clustering").
If the cardinalities $|Z|$ and $|X|$ are equal, we 
arrive at the fully detailed, trivial solution where the 
clusters $Z$ simply copy the original $X$. 
A formal 
solution to the information 
bottleneck problem is given in~\cite{bottleneck} and yields the 
following three self-consistent equations (with $\beta=1/T$),
\begin{equation}
\left\{
\begin{array}{ll}
p(z|x)&= \frac{p(z)}{Z(\beta,x)} e^{-\beta D_{KL}[p(y|x) || p(y|z)]} \\
p(z)& = \sum_x p(z|x)p(x)\\
p(y|z)&=\frac{1}{p(z)}\sum_x p(y|x) p(z|x) p(x) 
\end{array}
\right.
\label{eqt:sc}
\end{equation}
where $Z(\beta,x)$ is a normalization (partition) function and 
\nobreak{$D_{KL}[p||q] \equiv \sum_x p(x)\log \frac{p(x)}{q(x)}$} is the 
Kullback-Leibler divergence (also called the relative entropy). 
The first of these equations makes
clear that as one anneals to ground state, where $T\rightarrow 0$ and 
$\beta\rightarrow \infty$, 
the only solution is the hard clustering ($p(x|z)\in \{0,1\}$) limit.
These three equations naturally lend themselves 
to an iterative algorithm proposed in~\cite{bottleneck} 
which is provably convergent and finds a locally 
optimal solution. 

While in many applications a ``soft'' clustering might be of interest,  
for clarity we only consider the hard case in this paper:
each node is associated with one and only one module.
We use two different algorithms to find approximate solutions to the
information bottleneck problem. Both of them take a fixed $|Z|$
($|Z|<|X|$) as
input and output hard clustering assignments for every node.

The first algorithm ({\em self-consistent NIB}) uses $\beta$ as an 
annealing parameter that starts at low values and increases step by step.
At every given $\beta$ 
the locally optimal solution is computed by iterating over Equations (\ref{eqt:sc}). 
The solution for given $\beta$ is then taken as a starting 
point for the iterations with the next $\beta$.
The second algorithm ({\em agglomerative NIB}) uses an agglomerative approach~\cite{agg}. At every step
a pair of nodes is merged into a single node, where the pair is chosen
such as to maximize the relevant information $I(Y,Z)$. 
It thus reduces $|Z|$ by one at every step, and stops when the desired $|Z|$
is reached. 

\section{Diffusive Distributions Defined over Graphs}
\label{sec:diff}
We wish to find a representation of a network in which groups 
of nodes have been represented by  
{\em effective} nodes; 
we argue that a modular description of the network is most 
successful when relevant information about the network is preserved. 
Posed in this language, it is clear that the act of finding modules 
in a network is a type of clustering, and the appropriate clustering 
framework is one that preserves the information deemed relevant. 

Formulation of graph clustering in terms of the information 
bottleneck requires a joint distribution $p(y,x)$ to be defined on the graph,
where $x$ designates nodes and $y$ designates a relevance variable.
An appropriate distribution that captures structural information
about the network is the one defined by graph diffusion. 
The relevance random variable $y$ then ranges over the nodes,
as does $x$, and is defined by the node at which a random walker stands
at a given time $t$ if the random walker was standing at node $x$ at time $0$.
The conditional probability distribution $G_{ij}^t\equiv p^t(y_i|x_j)$ is
a Green's function describing propagation from node $j$ to node $i$. 
For discrete time  diffusion one can easily derive~\cite{Chung}
\begin{equation} 
G^t=[W T^{-1}]^t\;\;\mbox{, ($t\in\mathbb{N}$)}
\end{equation}
where $W$ is a symmetric weighted affitiny matrix of positive
entries and $T_{ij}\equiv\delta_{ij}\sum_l W_{il}\equiv \delta_{ij}k_i$. 
For a graph, with identically-weighted edges, 
$k_i$ is the conventional degree (the number of neighbors of node $i$),
and $W$
is the adjacency matrix ($W_{ij} = 1$ iff $i$ is adjacent to $j$).
Note that we here only consider connected graphs and, as defined,
this approach treats directed and undirected graphs identically.

In the continuous time limit 
\begin{equation}
G^t = e^{(W T^{-1}-1)t} = e^{-LT^{-1}t}\;\;\mbox{, ($t>0$)}
\end{equation}
where we defined $L\equiv T-W$, the graph Laplacian~\cite{merris}.
In the machine learning literature a ``graph kernel''~\cite{risi}
has been defined as 
\begin{equation}
G^t = e^{-Lt}.
\end{equation}
to learn from structured data. It corresponds to
a probability distribution associated with a different
diffusion rule assuming a
degree-dependent permeability at every node.
For comparison, we consider both of these joint distributions as possible
input to the information bottleneck algorithm.

The characteristic time scale $\tau$ of the system is given by the inverse
of the smallest non-zero eigenvalue of the diffusion operator exponent
($LT^{-1}$ or $L$ ). This time reflects the finite
system size and characterizes large-scale behaviors. For example, 
in one dimension on a bounded domain of size $\ell$, the smallest non-zero
eigenvalue of the Laplacian with diffusion constant $D$
is $\pi^2D/\ell^2$. 
For our algorithm we will thus choose $t=\tau$. 

To calculate the joint
probability distribution  $p(y,x)=p(y|x)p(x)$ from the conditional probability 
distribution  $G^{\tau} = p(y|x)$, we 
must specify a prior $p(x)$:  the distribution of random walkers at time 0.
Natural definitions include (i) a flat prior $p(x)={1}/{N}$, $N$ being the
total number of nodes and (ii) a prior corresponding to the steady state distribution 
associated with the diffusion operator: $p(x)={1}/{N}$ or
$p(x)=k_x/{\sum_xk_x}$, for $G^\tau=e^{-L\tau}$
or $G^{\tau}=e^{-LT^{-1}\tau}$, respectively, where $k_x$ is the degree of node $x$.

\section{Quantifying Modularity }
\subsection{Partition modularity -- quality of a partitioning} \label{sec:partmod}
Newman and Girvan 
~\cite{newman} propose a modularity, 
a ``measure of a particular division of a network'',
as $Q = \sum_i[ e_{ii} - (\sum_je_{ij})(\sum_ke_{ik})]$, 
where $e_{ij}$ is the fraction of all edges connecting 
module $i$ and module $j$. It can be interpreted as the difference between
the fraction of within-module edges and the expected fraction of 
within-module edges in an ensemble of networks created by randomizing 
all connections while holding constant the number of edges emanating from each
module. $Q$ should therefore go to $0$ for randomly connected networks, and 
tend to $1-1/|Z|$ for a perfectly modular network with $|Z|$ 
equally sized modules.
We herein refer to the measure $Q$ as {\em partition modularity} 
to distinguish it from {\em network modularity} which we define below 
based on information-theoretic quantities. 
Newman {\em et al.} also study the number of modules $|Z|_{\rm max}$ which maximizes $Q$ 
given a particular module discovery algorithm.

\subsection{Network modularity -- summarizability of network structure}
We here propose a new modularity measure $M$, a property of a given network rather than
of a given partitioning, which quantifies the extent to which a network
can be summarized in terms of modules.  

Every clustering solution $p(z|x)$
determines an normalized ``input information'' $0<I(Z,X)/H(X)<1$ between input variable 
$X$ and cluster assignment $Z$, and an ``output 
information'' $0<I(Z,Y)/I(X,Y)<1$ between cluster assignment and relevance variable $Y$.  
The {\em information curve}
is then
plotted as ${I(Z,Y)}/{I(X,Y)}$ vs. ${I(Z,X)}/{H(X)}$ for every 
solution of Equation (\ref{eqt:sc}) for every possible number of clusters~\cite{bottleneck}. 
An example is shown in Figure~\ref{infoplane}. 
The curve traced by minimizers of the functional Eqn. \ref{eqn:free_energy}, which
will not necessarily be computed by such approximating schemes
as the AIB, is provably convave. For perfectly random data,
which cannot be summarized, this curve lies along the diagonal $y=x$.
Consistent with these observations, we find that synthetic graphs
with high connectivity within defined modules, and low connectivity
between different modules exhibit larger area under the information
curve (data not shown). 
We thus define a new measure of {\em network modularity}: the area
under the information curve. In the soft clustering case the information curve
is continuous since solutions vary with every choice of $\beta\in[0,\infty)$.
In the hard clustering case, which we study here, the information curve is only
defined at discrete points corresponding to solutions for every possible
number of clusters $|Z|$. The area can then be calculated by linear 
interpolation. 
Information-theoretic bounds constrain the range of $M$ allowing comparison of networks 
of varying number of nodes and edges, and is a property of the network itself, rather than a given
partitioning.

\section{Tests on synthetic networks}
\newcommand{\erd}{Erd\"os~}
\subsection{Accuracy of the partitioning}
We here test how well various NIB implementations with different 
diffusion operators can reconstruct modules in a network
generated with a known modular structure. We also compare
our method to the ``edge-betweenness" algorithm recently proposed
in~\cite{newman} for the same purpose of finding modules or
``communities''. In~\cite{newman} the network is broken down
into isolated components by iteratively removing edges with highest
``betweenness'' 
(several definitions of edge-betweenness are tested in \cite{newman};
we here use the ``shortest path" betweenness, which was shown in \cite{newman}
to perform optimally). 

As in~\cite{newman}, we generate synthetic Erd\"os-like
graphs via Monte Carlo with 128 nodes each and average degree 8
(average total of 512 edges). We also demand that the graphs be connected by
rejecting generated graphs that have disconnected components.
We impose a structure of 4 modules with 32 nodes each by introducing two
different probabilities: $p_{\rm in}$ for edges inside modules and
$p_{\rm out}$ for edges between different modules.  
The level of noise in the graph is thus controlled by $p_{\rm out}$.  
The higher
$p_{\rm out}$, the harder it will be to recover the different modules.
We first generate networks with $p_{\rm out}=0$ and then increase
$p_{\rm out}$ while adjusting $p_{\rm in}$ such that the average degree
remains fixed. When $p_{\rm in}=p_{\rm out}$, all modular structure is
lost and we obtain a usual \erd graph.
We measure the accuracy of a proposed partitioning using the following 
computation. In principle any module proposed by the algorithm could
match any ``true'' module with an associated error. We therefore 
try every possible permutation of the 4 proposed modules matching
the 4 ``true'' modules, and consider the one permutation with the 
smallest total number
of incorrectly assigned nodes. We define {\em accuracy} as the total 
fraction of correctly assigned nodes. 

\begin{figure}[htb]
\begin{center}
\begin{tabular}{c}
\includegraphics[width=6 in]{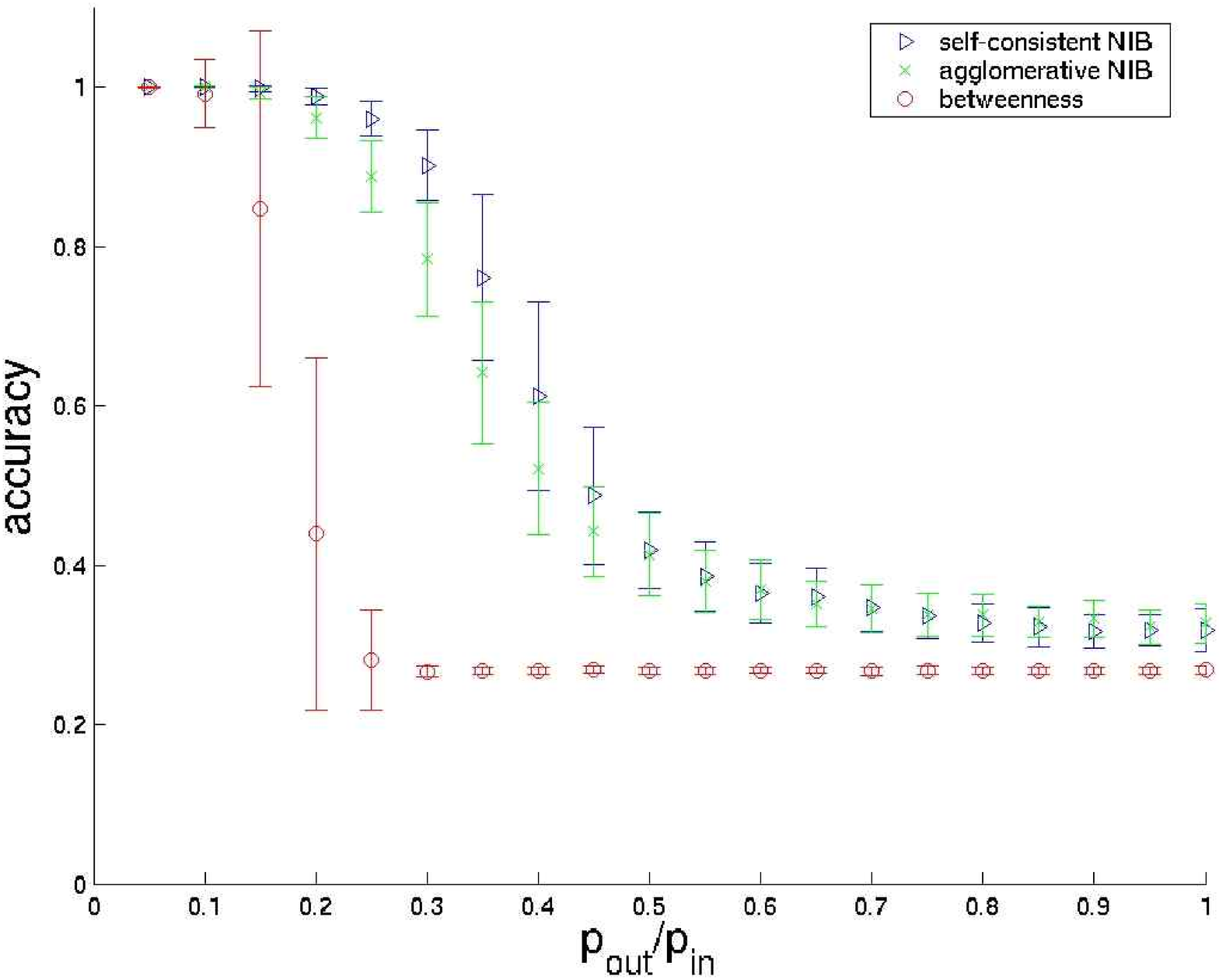}\\
{\bf (a)}\\
\includegraphics[width=6 in]{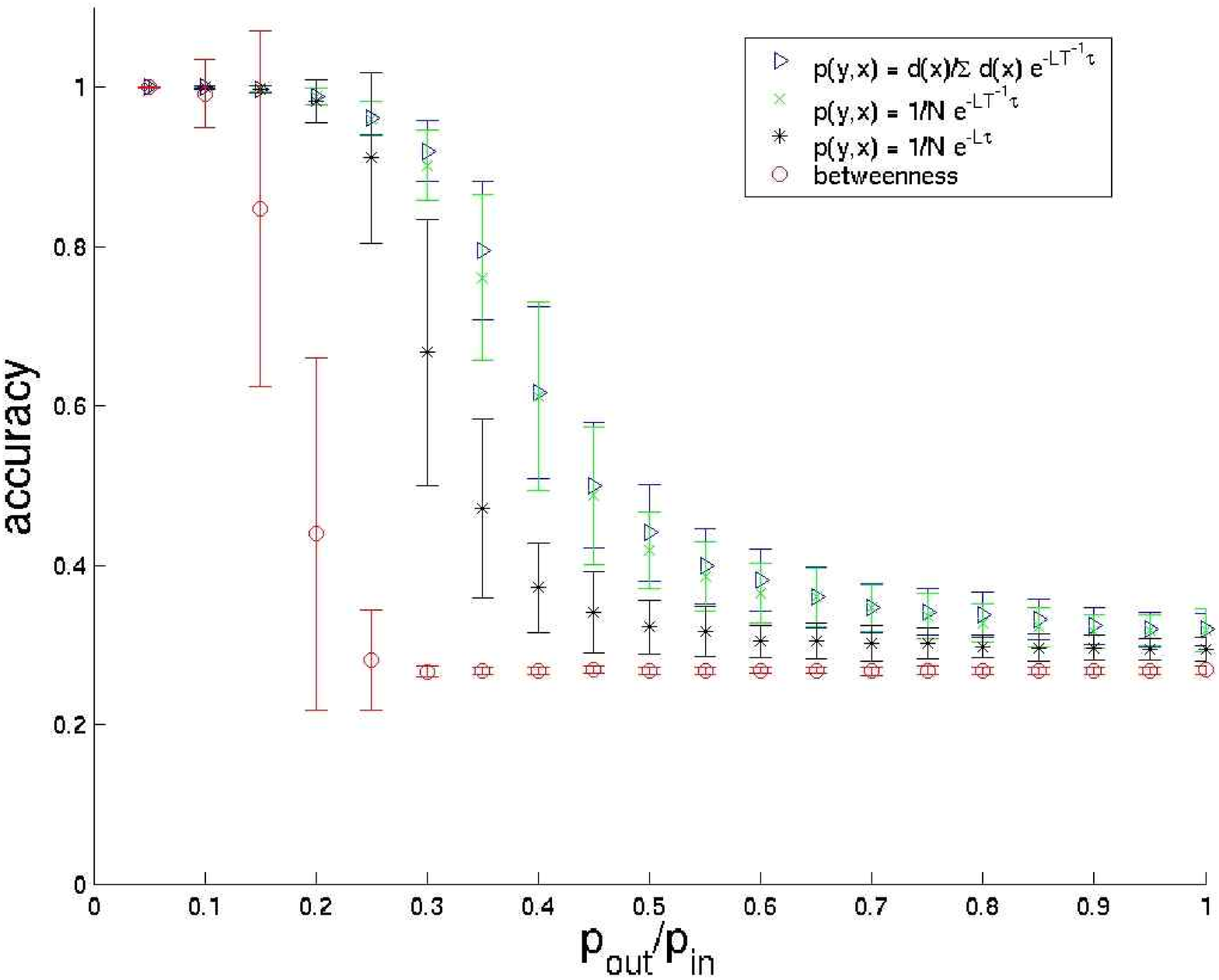}\\
{\bf (b)}
\end{tabular}
\caption{{\bf (a) Accuracy for different algorithms}. 
Measured is the accuracy 
of recovering modular structure in syntheticnetworks under varying 
noise levels. Every point represents an average over 100 networks, each with 128
nodes and an average degree of 8. Both NIB algorithms outperform the betweenness
algorithm.\\ 
{\bf (b) Accuracy for different diffusion operators}
Accuracy is measured in the same way as in (a), now using the self-consistent NIB
algorithm with various diffusion operators to define probability 
distributions $p(y,x)$ over nodes. For comparison the betweenness results are also
shown. The operator for physical diffusion $e^{-LT^{-1}t}$ outperforms the ``graph kernel''
diffusion operator proposed in the machine learning literature~\cite{risi}.
}
\label{modfig12}
\end{center}
\end{figure}

Figure~\ref{modfig12}a shows the accuracy of the recovered modules as a function
of ${p_{\rm out}}/{p_{\rm in}}$ for three different algorithms: 
self-consistent NIB, agglomerative NIB, and betweenness.
Both NIB algorithms use the physical diffusion operator $e^{-LT^{-1}t}$ and
a flat prior $1/N$ to define
a joint probability distribution. We observe that both NIB algorithms
are much more successful in recovering the modular structure than the
betweenness algorithm. A threshold noise level is achieved 
at around ${p_{\rm out}}/{p_{\rm in}}\approx 1/3$ for the NIB algorithms, and
around ${p_{\rm out}}/{p_{\rm in}}\approx 1/6$ for the betweenness
algorithm. The figure also shows that the self-consistent NIB in general
finds a better partitioning than the agglomerative NIB. 

Figure~\ref{modfig12}b shows the same measurements for self-consistent NIB algorithms
using different diffusion operators as explained in Section~\ref{sec:diff}. For
comparison the betweenness results are also plotted. 
Physical diffusion, 
defined by 
the continuous time limit, 
with an initial state $p(x)$ given by the equilibrium distribution
$p(x)\propto k_x$,
gives best performance.

\subsection{Finding the optimal number of modules}
In most real world problems the correct number of modules $|Z|$ present in the network
is unknown \emph{a priori}. It is therefore important to have an algorithm which not
only computes a good partitioning for a given $|Z|$ but also gives a good estimate for
$|Z|$ itself. 
To this end, we here make use of the partition modularity $Q$ as described in Section 
\ref{sec:partmod}.

\begin{figure}[htb]
\begin{center}
\begin{tabular}{c}
\includegraphics[width=6 in]{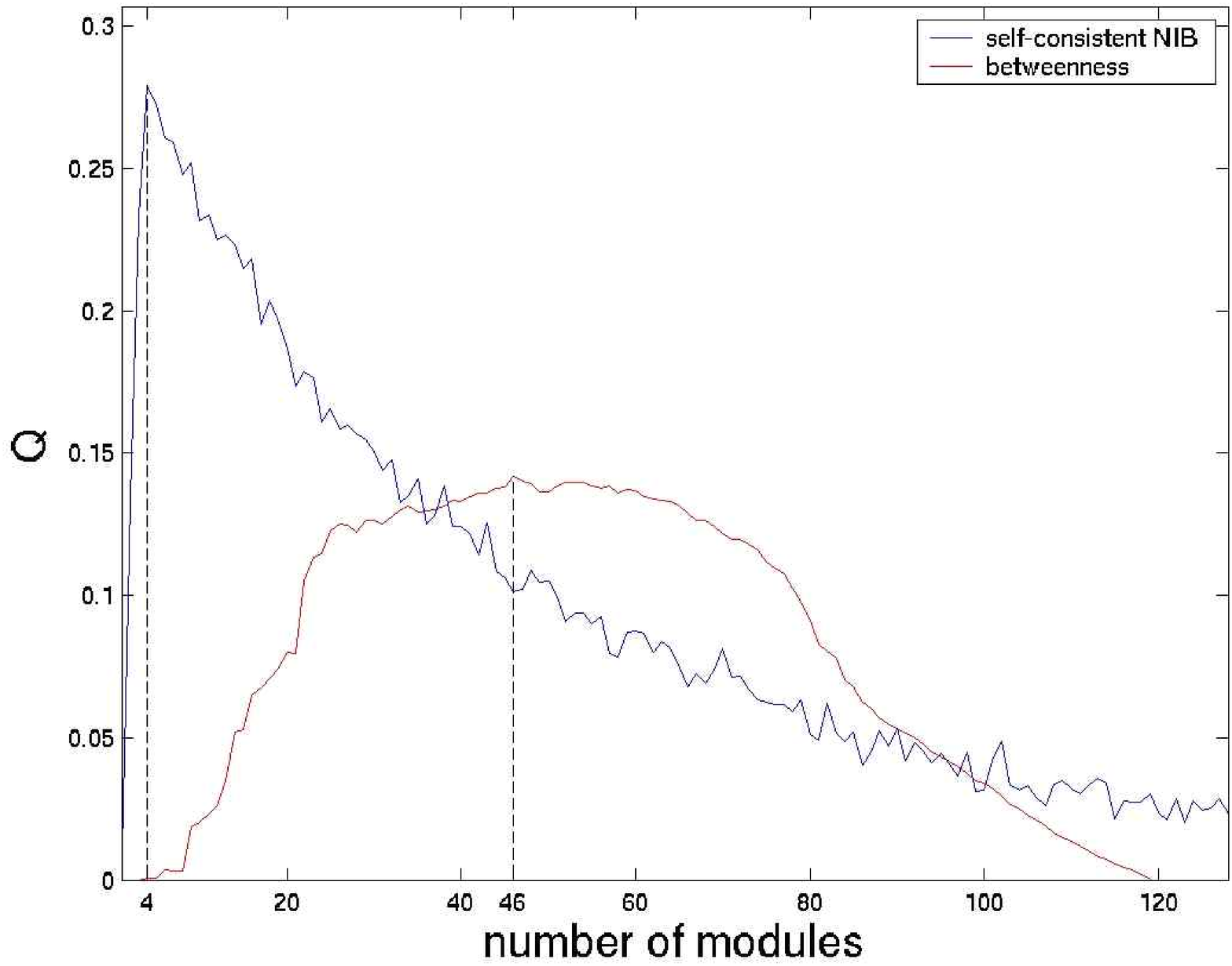}\\
{\bf (a)}\\
\includegraphics[width=6 in]{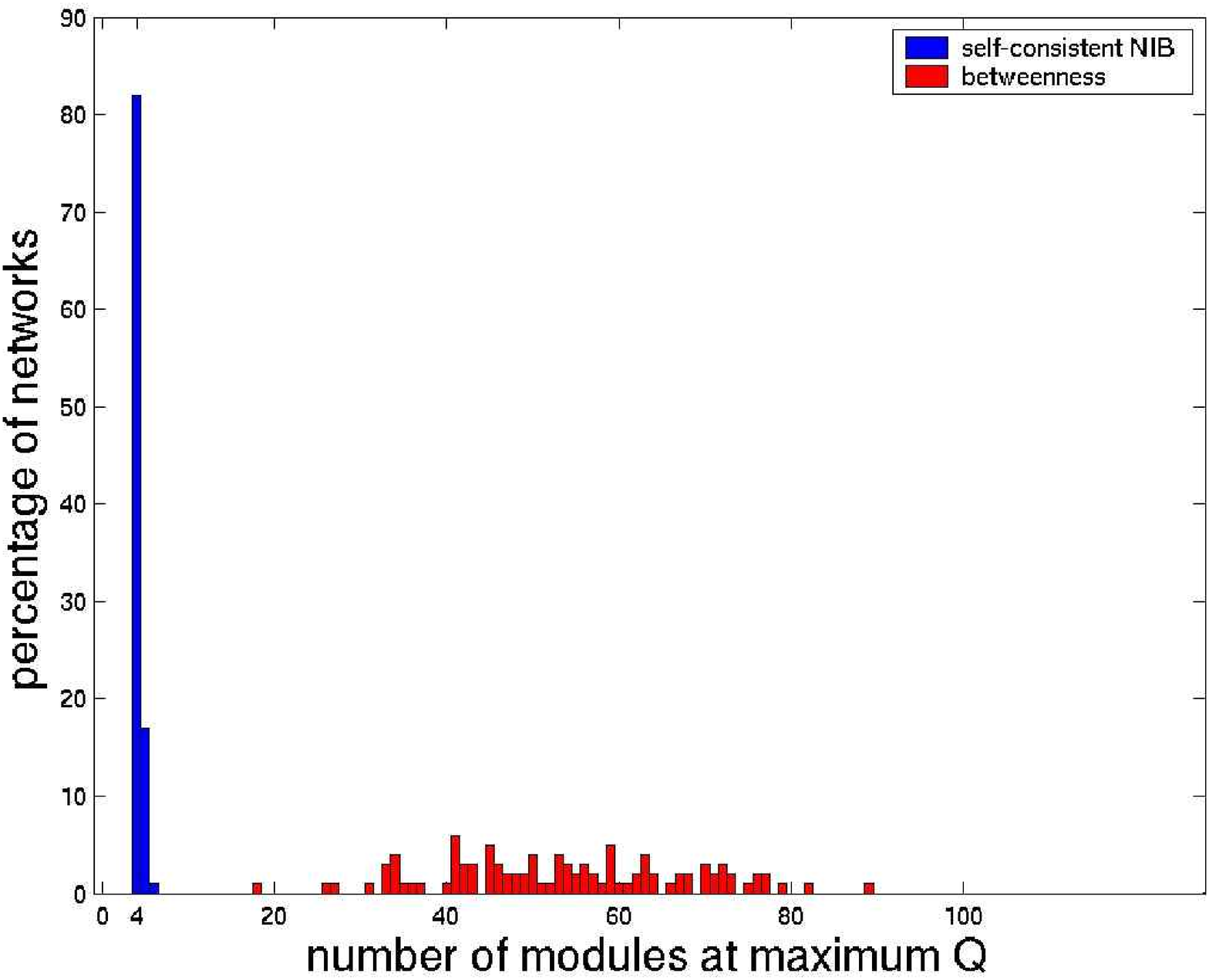}\\
{\bf (b)}
\end{tabular}
\caption{{\bf (a) $\mathbf{Q}$ vs. $\mathbf{|Z|}$}. The quality of the partitioning $Q$
is computed for the partitionings output by the self-consistent NIB
algorithm (with physical diffusion operator), and the betweenness algorithm,
for every given number of modules $|Z|$. While the NIB algorithm correctly
determines $|Z|=4$ as the best number of modules (at maximum $Q$), 
$Q$ calculated by the betweenness algorithm peaks at $|Z|=46$.  \\ 
{\bf (b) Histogram of $|Z|_{\rm max}$ for 100 different networks. }
For 82\% of the networks, the NIB algorithm is able to find the correct
number of modules $|Z|_{\rm max}=4$, and comes close to it ($|Z|_{\rm max}=5$ or $6$)
in all other cases. The betweenness algorithm gives $|Z|_{\rm max}$ between
$18$ and $89$, far from the correct value for all networks. 
}
\label{modfig34}
\end{center}
\end{figure}

We again consider synthetic connected networks of 128 nodes and 
average degree 8 as in the previous section. However, we fix the noise level
to a value of ${p_{\rm out}}/{p_{\rm in}} = 0.3$ which was shown to be a critical
level for these networks. 
We run the self-consistent NIB and the betweenness algorithms for every possible
number of modules $|Z|=1,2,\dots ,128$ and compute $Q$ for the proposed 
partitionings. Figure~\ref{modfig34}a shows $Q$ as a function of $|Z|$ for a 
typical run. While for the NIB algorithm $Q$ sharply peaks at the correct value
of $|Z|_{\rm max}=4$, $Q$ calculated by the betweenness algorithm attains 
its maximum at $|Z|_{\rm max}=46$ and does not show a particular signal at $|Z|=4$. 
Figure~\ref{modfig34}b shows a histogram of $|Z|_{\rm max}$ for 100 generated
networks. The NIB algorithm successfully identifies $|Z|_{\rm max}=4$ for 82\% of the networks,  
while the betweenness algorithm calculates $|Z|_{\rm max}$ lying between $18$ and $89$,
notably far from the correct value for any network. 
These experiments suggest that the NIB algorithm performs well 
both in accurately assigning nodes to modules and 
in revealing the optimal scale for partitioning.

\section{Applications}
\subsection{Collaboration Network}
Having validated NIB on a toy model of modular networks, we next apply our algorithm 
to two examples of naturally-occurring networks. In the first example, we construct 
a collaboration network from the 2004 APS March Meeting, where
this algorithm was first presented, and in the second example we 
construct a symmetric version of the \emph{E. coli} genetic 
regulatory network. 

\begin{figure}[htb]
\begin{center}
\begin{tabular}{c}
\includegraphics[width=6 in]{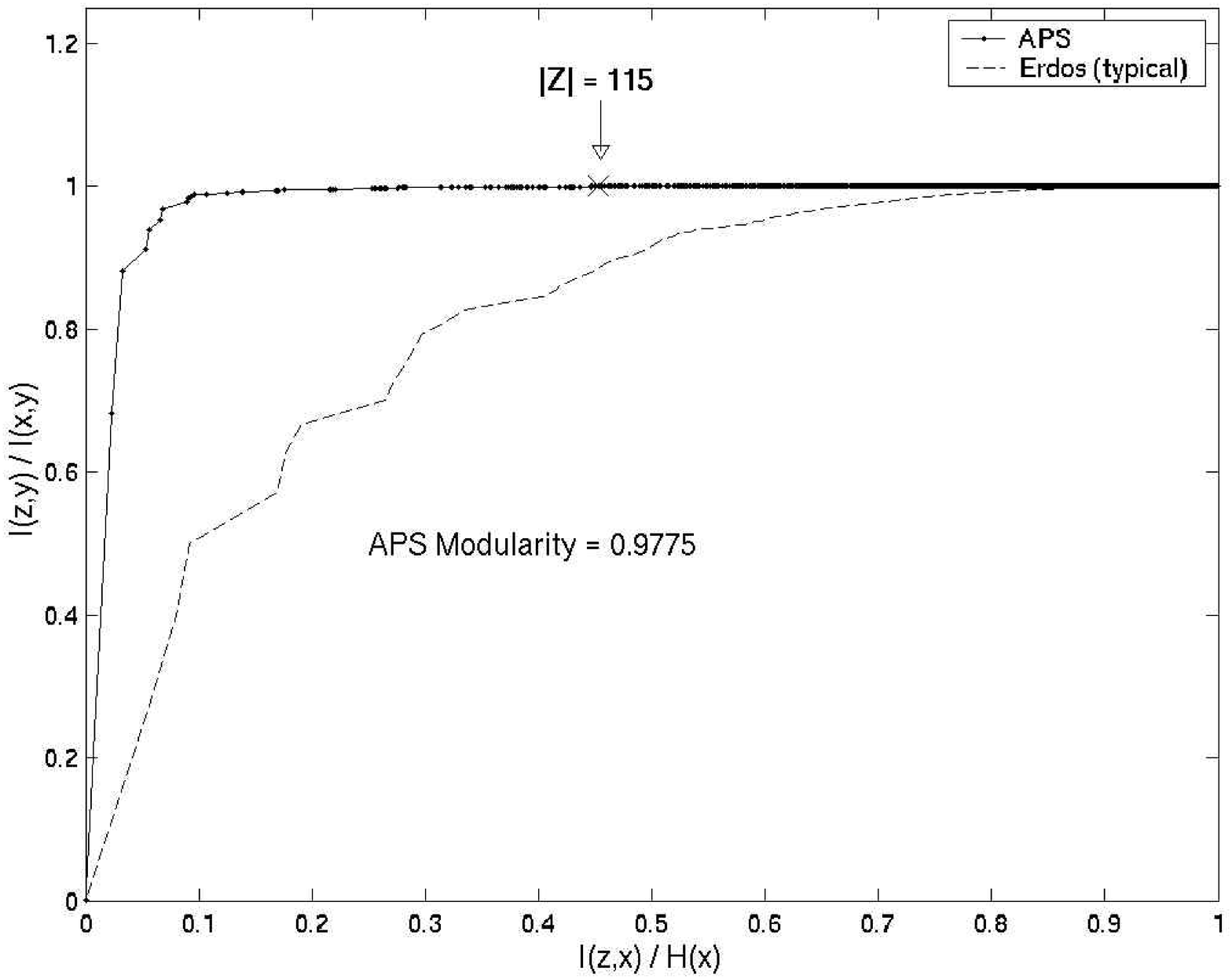}\\
{\bf (a)}\\
\includegraphics[width=6 in]{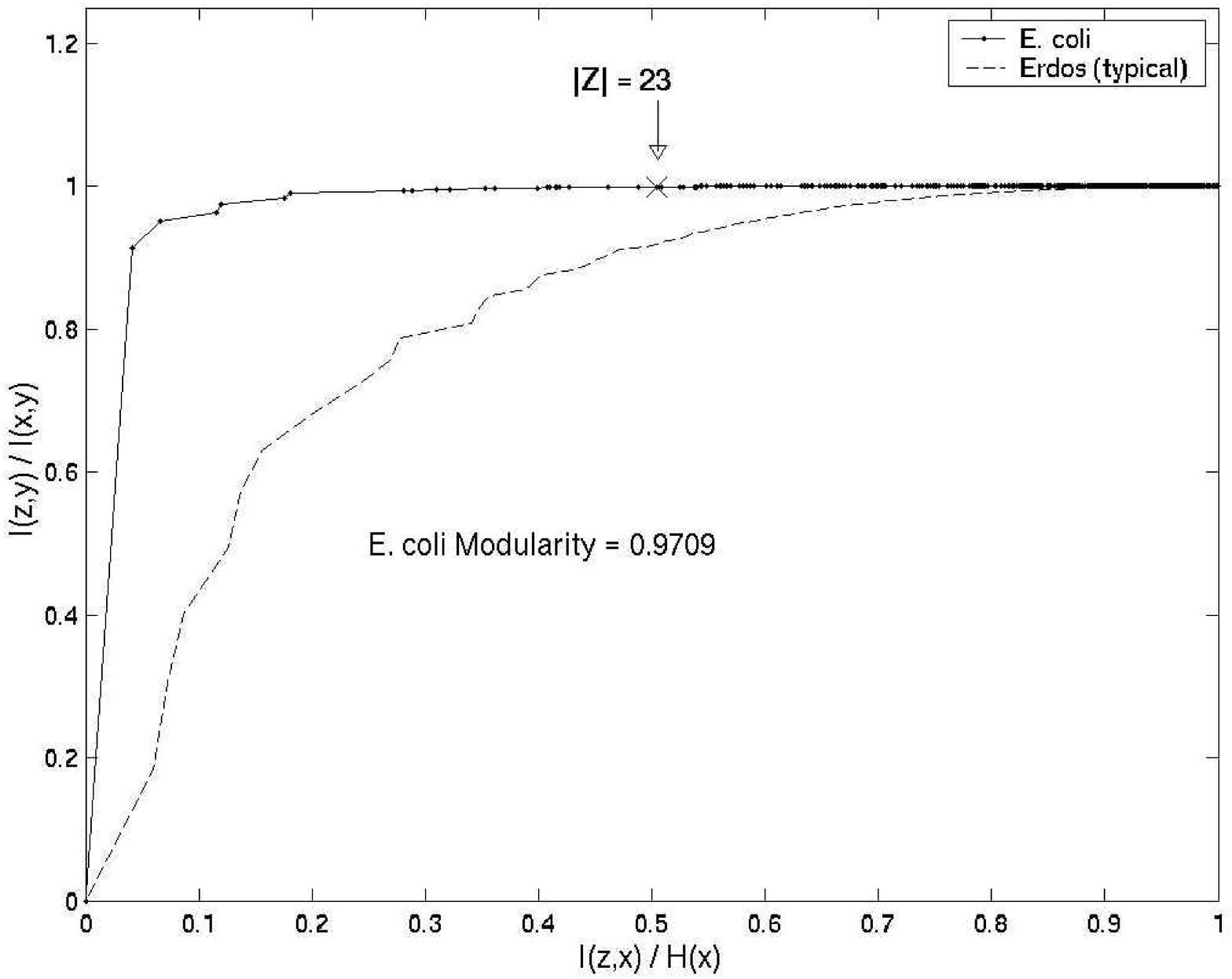}\\
 {\bf (b)}
\end{tabular}
\caption{{\bf (a) APS Network Modularity}. Information plane for the collaboration network 
obtained from the 2004 APS March Meeting (largest component consists of $5603$ 
authors and $19761$ edges). We use the agglomerative algorithm with the diffusion 
operator $e^{-LT^{-1}t}$. Network modularity for this graph, defined as the area 
under the curve is $0.9775$. Comparison is made with a typical information curve 
obtained from an Erd\"os graph. The optimal number of modules as defined by the the Newman and Girvan measure is at $|Z|=115$.\\
{\bf (b) \emph{E. coli} Network Modularity}. Information plane for the \emph{E. coli} 
genetic regulatory network (largest component $328$ nodes and $456$ edges). 
We use the agglomerative algorithm with the diffusion operator $e^{-LT^{-1}t}$. Network 
modularity for this graph, defined as the area under the curve is $0.9709$. Comparison is made with a typical information curve obtained from an Erd\"os graph. The optimal number of modules as defined by the the Newman and Girvan measure is at $|Z|=23$.}
\label{infoplane}
\end{center}
\end{figure}

Vertices of the collaboration network represent authors from all talks at the March Meeting; 
edges represent coauthorship. 
The largest component of the resulting graph consists of $5603$ 
vertices with $19761$ edges. 
Network information bottleneck using the agglomerative algorithm and 
the physical diffusion operator (as defined in Section \ref{sec:diff}
with its corresponding equilibrium distribution) 
reveals that this large network is highly modular ($m=0.9775$, see 
Figure \ref{infoplane}a). For comparison, we also show the information curve for a typical Erd\"os 
network, which is clearly less modular.  
Such a high value of modularity implies that 
the authors of this component of the network are ``easily" compressed or 
combined into larger clusters of authors. In the light of this fact, we study
what the clusters of authors reveal about the collaboration 
network. For example, authors may group themselves according to topics 
or subject matters of the talks; alternatively, author modules may be 
more indicative of the authors' affiliations or even geographical location. 

To begin to approach these types of questions we may choose to look at the author 
groupings given by NIB at a particular number of clusters. While we emphasize 
that network modularity is a measure over all scales or all numbers of clusters, it is illustrative in this case also to examine the clustering at a particular scale. 
For the APS network, such an analysis yields the optimal number of 
modules, $|Z|_{\rm max}=115$, for this network (see Figure \ref{infoplane}a). 

\begin{figure}[htb]
\begin{center}
\includegraphics[width=3 in]{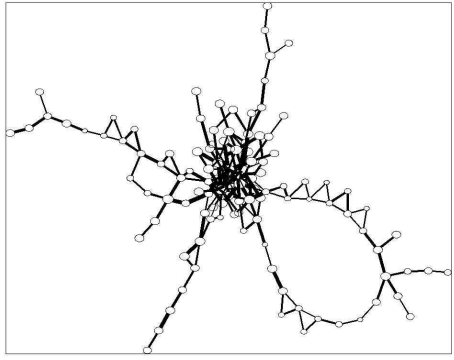}
\caption{Adjacency network of the $115$ modules of the APS network. Nodes represent modules 
where the size of the drawn ellipse is proportional to the number of authors 
in the module. Edges between modules represent collaborations between authors in 
different modules, where the thickness of the drawn lines is proportional to the 
number of these inter-module collaborations. The module-network reveals a structure 
with highly dense center of modules, three branching linear chains of modules, and one cycle of modules.}
\label{aps_adj}
\end{center}
\end{figure}

In Figure \ref{aps_adj} we plot the $115$ modules and their connections 
where each ellipse represents one module and edges between ellipses represent 
inter-modular connections. The sizes of the ellipses and the thickness 
of the edges are proportional to the log of the number of authors 
in a module and the log of the number of inter-modular connections 
between modules, respectively.
We note the provocative structure revealed in the figure with a large 
center of highly connected modules (including two of the largest modules), 
three more or less branching, linear chains of modules, and one
large $18$-node cycle of modules. 

\begin{figure}[htb]
\begin{center}
\includegraphics[width=3 in]{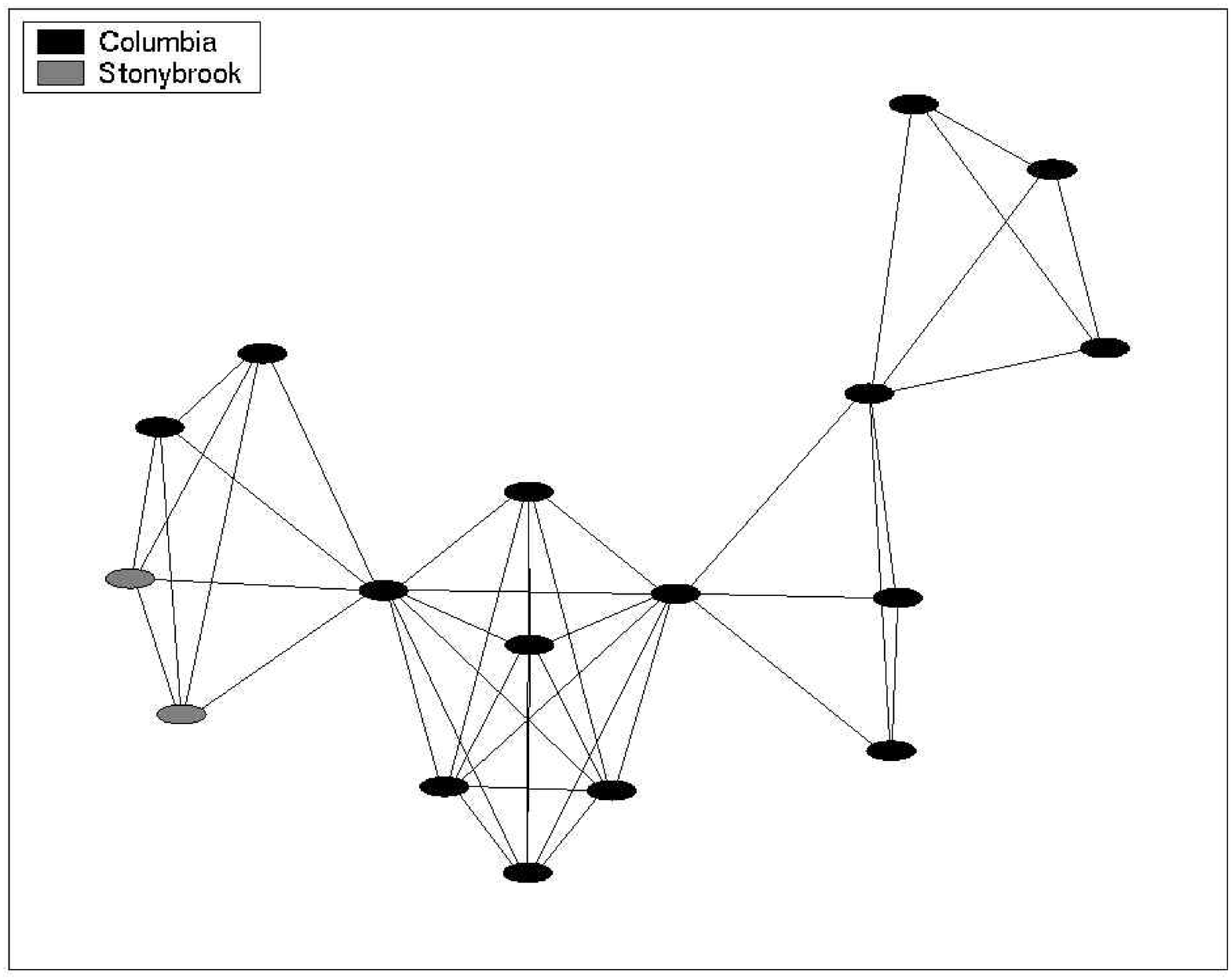}
\caption{One of the $115$ modules of the APS network. Nodes represent authors and edges 
represent collaborations. Of the $17$ authors in the module, $15$ are Columbia-University 
affiliated and two are affiliated with Stony Brook University.}
\label{col_mod}
\end{center}
\end{figure}

Closer inspection of a single module
demonstrates that for many of the modules, institutional affiliation, 
and even geography, play a large role in determining collaborations. 
In Figure \ref{col_mod} a single $17$-node module is plotted where 
each node now represents an author and edges represent author collaborations.  
We see that $15$ of the $17$ authors are affiliated with Columbia University;
the remaining two authors are affiliated with Stony Brook, and notably, are 
adjacent (indicating coauthorship) to each other. The finding that the modules in this collaboration 
network are somewhat related to institutional affiliations and 
geography is supported by similar results found in other 
physics collaboration networks previously studied using different techniques \cite{newman}. 

Another possible annotation for this module to consider is that of the 
APS divisions and topical groups, since each author is associated with at 
least one talk and each talk is listed under one or more of these APS 
categories. However, the $14$ APS divisions and $10$ topical groups appear to be 
too broad and have too much overlap to clearly define a module. 
For example, the Columbia University module includes talks under 
the categories of Polymer, Condensed Matter, Material, and Chemical Physics. 
On the other hand, the module is essentially representative of researchers 
at the Columbia University Materials Research Science and 
Engineering Center (MRSEC) and in particular those interested in the 
synthesis of complex metal oxide nanocrystals. There is thus both topical 
and institutional information retained in the modules. 

\begin{figure}[htb]
\begin{center}
\includegraphics[width=3 in]{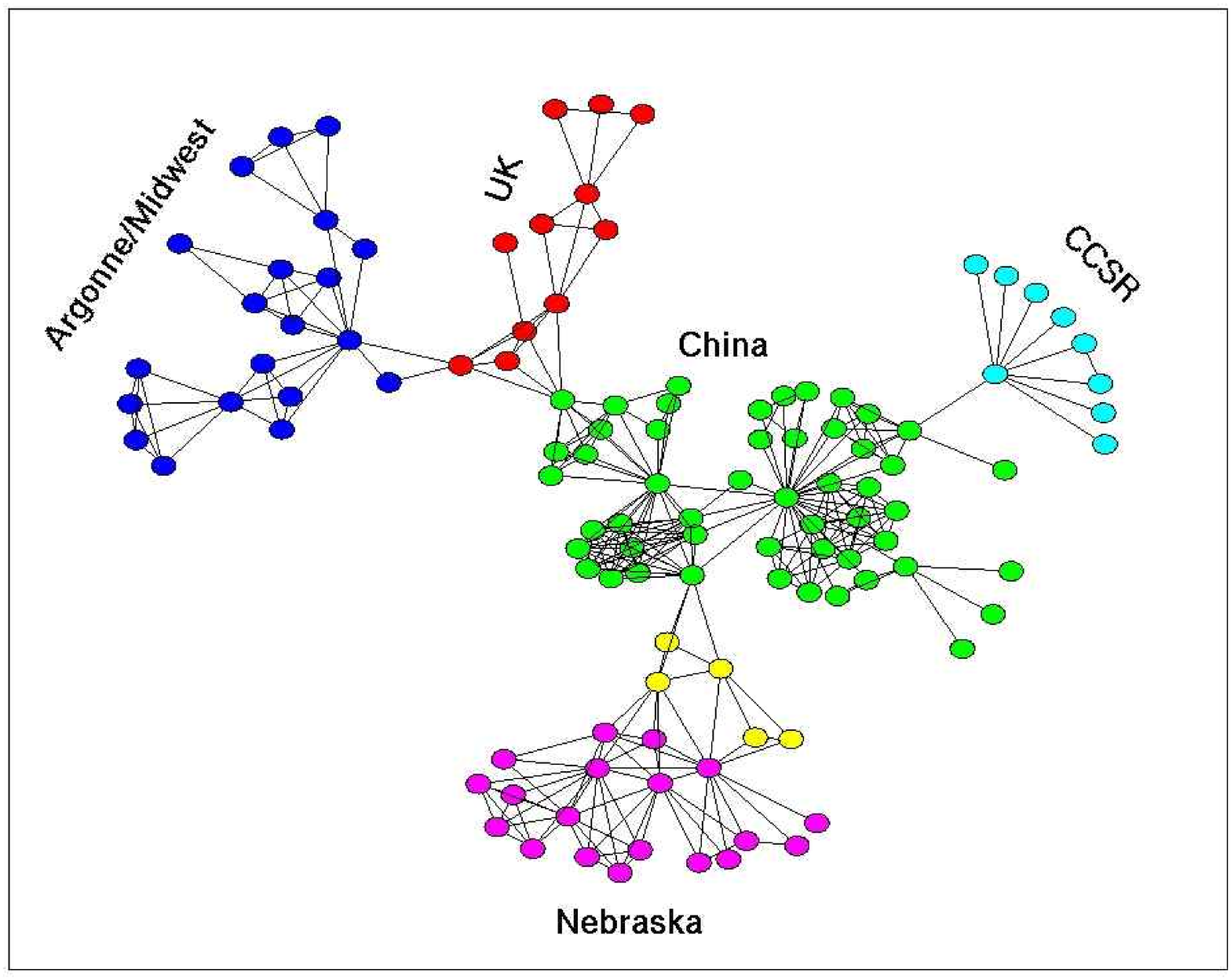}
\caption{Six modules corresponding to the uppermost branched linear chain of modules depicted 
in Figure \ref{aps_adj}. Colors denote modules as defined by the network information 
bottleneck algorithm. Again the modules roughly correspond to institutional affiliations. 
Over $50\%$ of the blue nodes have one or more affiliations with the institutions based 
in and around Chicago (Argonne National Laboratory, University of Illinois at Chicago 
and University of Notre Dame). $70\%$ of the red nodes are in England, and $75\%$ of the 
green nodes are in China, mostly at the Institute of Chemistry Chinese Academy of 
Sciences, and all of the cyan nodes are at the Center of Complex Systems Research 
in Illinois. Both the yellow and magenta modules are mostly affiliated with the University of Nebraska.}
\label{6_mods}
\end{center}
\end{figure}

It is also revealing to examine the affiliations of multiple connected modules. 
For example, Figure \ref{6_mods}  
plots the uppermost branching linear chain of Figure \ref{aps_adj}.
Here, color denotes module assignment as given by NIB.
Most of these modules also have clear institutional affiliations. For example, 
everyone in the cyan module is at the Center of Complex Systems Research (CCSR) 
in Illinois; close to $80\%$ of the large green module is in China, mostly at 
the Institute of Chemistry Chinese Academy of Sciences (ICCAS); and $70\%$ of 
the red module is in England. The blue module is slightly more diffuse, though 
an institutional affiliation is also apparent here; over $50\%$ of 
the authors are affiliated with one of three institutions near Chicago
(Argonne National Labs, University of Illinois at Chicago, and University of Notre Dame). 
The yellow and magenta modules are also overwhelmingly associated with 
the University of Nebraska, though interestingly our algorithm 
separates these two modules at this partitioning. In general, one does not
anticipate that the optimal number of clusters in a given network will give
the most natural partitioning at all scales and over all resulting modules.

\subsection{Biological Network}

The notion of modularity has been central in the study of a variety of biological 
networks including metabolic \cite{barabasi}, protein \cite{maslov, party_hub}, 
and genetic \cite{modular:murray:nature, alon02b} networks. Certainly most 
biologists agree that the various networks operating within and between cells 
have a modular structure, though what they mean by ``modular" 
can vary greatly \cite{wagner}. 

\begin{figure}[htb]
\begin{center}
\includegraphics[width=3 in]{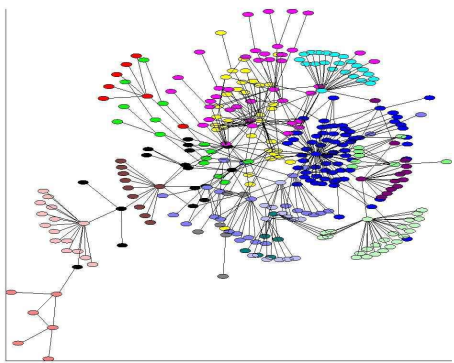}
\caption{{\bf \emph{E. coli} gene regulatory network}. 
Largest component of the symmetric version of the \emph{E. coli} genetic regulatory 
network. Colors denoted modules identified by NIB. 
}
\label{coli_full}
\end{center}
\end{figure}

NIB allows us to investigate quantitatively and in detail to what 
extent naturally-occurring biological networks are modular. For example, 
Figure \ref{coli_full} depicts the undirected form of the largest component 
of the \emph{E. coli} genetic regulatory network described previously in 
\cite{alon02b} and \cite{regulondb}. The network consists of $328$ vertices 
and $456$ edges and its modularity is depicted by the curve one traces in 
the information plane as the network is clustered using the network 
information bottleneck (see Figure \ref{infoplane}b).

\begin{figure}[htb]
\begin{center}
\includegraphics[width=3 in]{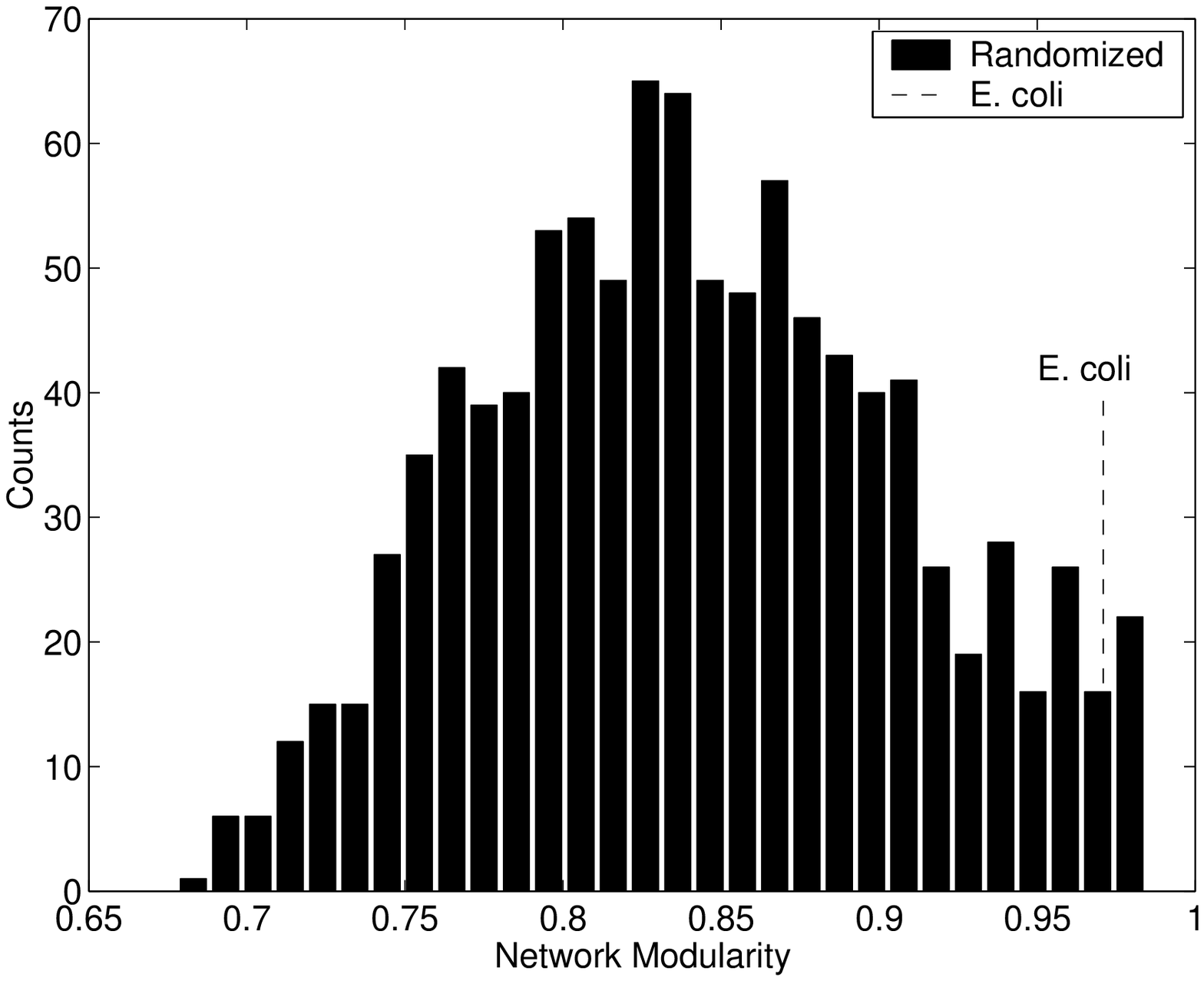}
\caption{A histogram of network modularity, defined by the area underneath the curve in 
the information plane resulting from network compression, for $1000$ realizations of 
the variation of the configuration model. Note that the true \emph{E. coli} network is 
more modular than the typical network resulting from the variation of the configuration model.}
\label{coli_rand}
\end{center}
\end{figure}

To establish whether the modularity of the network should be considered low, 
high, or moderate, we employ 
an ansatz popular in several reserach communities
in which a distribution of networks is created by holding the in-, out-, and 
self-degree of each node constant but randomizing the connectivity 
of the graph, changing which nodes are connected to which neighbors 
\cite{alon02b,maslov,ziv:matstat,HandL,snijders,ecology}. 
The randomization, a 
variant of the configuration model~\cite{newman_review}, produces a distribution of networks from 
which we sample and then measure the network modularity. The histogram 
in Figure \ref{coli_rand} shows that {\em E. coli}'s modularity
is higher relative to this ensemble.


\section{Conclusions and Extensions}
We have presented a principled, quantitative, parameter-free,
information-theoretic definition of network modularity, as well as an
algorithm for discovering modules of a network. Network modularity is
a dimensionless number between $0$ and $1$ and is a property of a
given network over all scales, rather than of a given partitioning
with a given number of modules. The measure is applicable to any
network, including those with weighted edges. We validate the
effectiveness of our algorithm in identifying the correct modules and
in finding the true number of modules on synthetic, Monte Carlo generated,
Erd\"os-like, modular networks. Finally, application to two real-world
networks, a ``social" network of physics collaborations and a
biological network of gene interactions, is demonstrated.

Network modularity, the area under the curve in the information plane,
is but one relevant statistic that we may retrieve from the
information curve. Certainly other useful statistics may be
culled. For example, the optimal information curve will always be
concave~\cite{slonimthesis} and its slope will decrease
monotonically. The point at which the slope equals $1$ is uniquely
determined for each network and can be described as the point after
which clustering further results in a greater loss in relative
relevant information than gain in relative compression (that is,
$\delta \frac{I(Z,Y)} {I(X,Y)} = \delta \frac{I(Z,X)}{H(X)}$).  This
{\em break-even point} is the point at which one can gain further
(normalized) simplicity only by losing an equivalent (normalized)
fidelity. Numerical experiments and investigating the utility of this
measure are currently in progress.

Diffusive distributions 
are but one general
class of distributions on a network. A natural
generalization of these ideas is to describe other distributions on a
network for which a particular function, energy, or origin
is known, and on which some particular degree of freedom (such as chemical
concentration or genetic expression as a function of time) may be
defined.

Finally, we note that while the information bottleneck is a
prescription for finding the 
highest-fidelity summary of a system at a given simplicity, 
algorithms for determining
network community structure  
are usually motivated 
by various definitions of normalized min-cuts
\cite{Weiss,Malik,Ng,Jordan03}. Our results, particularly for
the synthetic graphs with prescribed modular structure,
 demonstrate that
\emph{information} modularity implies \emph{edge} modularity, an
unexpected
finding which motivates further numerical and analytic investigations
in progress
regarding this relationship.

\subsubsection*{Acknowledgments}
It is a pleasure to acknowledge
Mark Newman,
Noam Slonim,
Christina Leslie,
Risi Kondor,
Ilya Nemenman,
and 
Susanne Still
for many useful conversations on the information
bottleneck and modularity.
This work was supported by
NSF ECS-0425850,
NSF DMS-9810750, and
NIH GM036277.

\bibliographystyle{plain}
\bibliography{infoarxiv}

\end{document}
\end